INDICATIONS OF R-MODE OSCILLATIONS IN SOHO/MDI SOLAR RADIUS MEASUREMENTS

P.A. Sturrock[1], R. Bush[2], D.O. Gough[3,2], J.D. Scargle[4]

[1] Center for Space Science and Astrophysics, and Kavli Institute for Particle Astrophysics and Cosmology, Stanford University, Stanford, CA 94305-4060, USA; sturrock@stanford.edu
[2] W. W. Hansen Experimental Physics Laboratory, Stanford University, Stanford, CA 94305, USA
[3] Institute of Astronomy and Department of Applied Mathematics and Theoretical Physics, University of Cambridge, Madingley Road, Cambridge CB3 0HA, UK
[4] NASA/Ames Research Center, MS 245-3, Moffett Field, CA 94035, USA

ABSTRACT
Analysis of solar radius measurements acquired by the Michelson Doppler Imager on the SOHO spacecraft supports previously reported evidence of solar internal r-mode oscillations in Mt Wilson radius data and in $^{90}$Sr beta-decay data. The frequencies of these oscillations are compatible with oscillations in a putative inner tachocline that separates a slowly rotating core from the radiative envelope.

1 . INTRODUCTION

A synoptic program of observations of the Sun by the 150-foot telescope at the Mt Wilson Observatory commenced in 1968 and continued until January 22, 2013. Radius measurements were obtained from the daily observations by a code that is not part of the main data reduction pipeline. The procedure used for determining the radius from the Mt Wilson observations has been described by Levebvre et al. (2004). Power spectrum analysis of the radius measurements for the time interval 1968.670 to 1997.965 yielded evidence of r-mode oscillations in the solar interior (Sturrock & Bertello 2010). The frequencies of these oscillations match those to be expected in a region where the sidereal rotation rate is 12.08 year$^{-1}$.

This rotation rate is slower than that of either the radiative zone or the convection zone (Schou et al. 1998). Super-Kamiokande data yield evidence that the solar neutrino flux varies with a periodicity of 9.43 year$^{-1}$ (Sturrock & Scargle 2006), which would correspond to a sidereal rotation rate of 10.43 year$^{-1}$. Since r-mode oscillations are likely to occur in a region where there is a gradient in angular velocity by a Kelvin-Helmholtz type of instability (Papaloizou & Pringle 1978), similar to that recently proposed by Zaqarashvili et al. (2010), it seems possible that the r-mode oscillations apparent in these solar radius measurements may have their origin in an "inner tachocline" that separates the core from the radiative zone.

The Michelson-Doppler Imager (MDI) on NASA's Solar and Heliospheric Observatory (SOHO) was in operation from 1997.843 to 2011.282. Although the primary goal was helioseismology, the mission also yielded precise measurements of the solar diameter. The purpose of this article is to report a power-spectrum analysis of the MDI radius (semi-diameter) measurements, the goal of which was to seek confirmatory evidence for the r-mode oscillations found in the Mt Wilson data.

Emilio et al. (2000), Kuhn et al. (2004) and Bush et al. (2010) have shown how the solar radius and its variation can be accurately inferred from the Michaelson Doppler Imager (MDI) image time series. These analyses explored several algorithms for obtaining the radius. In general, the solar radius is obtained from each image in several solar position-angle sectors from the apparent solar limb-darkening function as seen through the MDI optics. Further details may be found in the cited articles. As we see from Figure 4 of Levebvre et al. (2004), there is good agreement between radius measurements made from Mt Wilson data and from MDI data.

As will be discussed further in Section 5, a further impetus for searching for evidence of r-mode oscillations in MDI data has been the discovery that $^{90}$Sr beta-decay data acquired at the Lomonosov Moscow State University (LMSU; Parkhomov 2010a, 2010b, 2011) also appear to yield evidence of solar r-mode oscillations (Sturrock et al., 2012, 2013). MDI data may therefore prove to have significance not only for the structure of the solar interior, but also for nuclear physics.

2 . R-MODE OSCILLATIONS

To a good approximation, a star rotating slowly like the Sun may be considered to be spherical. Then the r-mode eigenfunctions are separable in radial and horizontal (angular) coordinates. In a star rotating uniformly with angular frequency $\nu_R$, the angular dependence is harmonic (as indeed it is also if the angular frequency of the star varies only with radius). In a frame of reference rotating with the star, r modes propagate retrograde and have cyclic frequencies that are well approximated by

$$\nu(l,m) = \frac{2m\nu_R}{l(l+1)}, \quad l \geq 1, \quad 1 \leq m \leq l, \tag{1}$$

where $l$ and $m$ are the degree and order of the associated spherical harmonic. The oscillatory motion is nearly horizontal, which is why the oscillations do not, to this order of approximation, sense the pressure and density stratification of the star. Therefore they can exist also in a vertical shear, being concentrated within some thin spherical shell rotating with mean frequency $\nu_R$.

We here interpret the frequency given by Equation (1) as the oscillation frequency caused by the interaction of a (retrograde) r mode with a non-axisymmetric structure, such as a magnetic-field configuration, that rotates with the solar interior at angular frequency $\nu_R$. That field is likely to be the relic of a primordial field, the dominant component of which is today dipolar, about an axis that is likely to be inclined with respect to the axis of rotation. We therefore expect the interaction to be strongest for r modes with $m = 1$. We envisage that the interaction causes a periodic modulation of the Sun's diameter with no detectable angular dependence. Therefore, any measurement of the modulation frequency would have the same value whether the diameter were observed from an inertial frame of reference or from Earth.

3 . POWER-SPECTRUM ANALYSIS

MDI observations yielded 448,168 MDI radius measurements acquired over the time interval 1997.843 to 2011.282. Since we are here interested only in frequencies less than a few cycles per year, we have reformatted the data into 1300 bins, each containing 344 or 345 entries with mean duration 3.776 days. Figure 1 shows the residuals (after taking account of instrumental effects) of the binned measurements.

We have carried out a power-spectrum analysis of these data by means of a likelihood procedure (Sturrock et al. 2005a) that is an extension (providing amplitudes and phases as well as powers) of the Lomb-Scargle periodogram procedure (Lomb 1976, Scargle 1982). As in the Lomb-Scargle procedure, the power at a given frequency is the square of the amplitude of the best-fit sine wave when measurements are normalized so that the variance is unity. The power spectrum contains peaks at frequencies 0.08 year$^{-1}$ and 0.15 year$^{-1}$, which we suspect arise from the sunspot cycle. Since the length of the dataset is only 13 years, we have chosen to remove these oscillations. This was achieved by determining the amplitude and phase of an oscillation, constructing a sine wave with the same amplitude and phase, and subtracting it from the data. We also removed the annual variation (by the same procedure) since this is presumably due to observational conditions that have no solar significance. The resulting power spectrum is shown in Figure 2.

We now determine whether there is a frequency that, if interpreted as a sidereal rotation frequency, leads to a group of r-mode oscillations of azimuthal order $m = 1$ that show up in the power spectrum. To this end, we follow the procedure previously used in the analyses of Mt Wilson radius data and LMSU beta-decay data.

We consider possible r modes with degrees up to 10 and with $m = 1$, as in both of the previous analyses. We then combine the powers of the oscillations identified with r-mode frequencies in the power spectrum by forming the "combined power statistic" (*CPS*; Sturrock et al. 2005b), defined as follows:

If $U$ is the sum of the powers at $n$ specified frequencies,

$$U = S_1 + S_2 + \ldots S_n , \qquad (2)$$

then the CPS is defined by

$$G(U) = U - \ln\left(1 + U + \frac{1}{2}U^2 + \ldots + \frac{1}{(n-1)!}U^{n-1}\right) . \qquad (3)$$

This statistic has the property that if each power arises from an exponential distribution (as is appropriate if the power is drawn from a time series dominated by normally distributed random noise; Scargle 1982), then $G$ also conforms to an exponential distribution.

This statistic is shown, as a function of the sidereal rotation frequency $\nu_R$, in Figure 3. We see that the function has a peak close to the same frequency (12.08 year$^{-1}$) that we found in our analysis of Mt Wilson data. Figure 4 shows the low-frequency range (0 – 5 year$^{-1}$) of the power spectrum (having removed oscillations with frequencies 0.08 year$^{-1}$, 0.15 year$^{-1}$ and 1.00 year$^{-1}$). We see that peaks corresponding to r-mode oscillations with $m = 1$, $l = 3, 4, 6$ and $10$ can be identified; the rms deviation $\sigma$ of their frequencies from those given by Equation (1) with $\nu_R = 12.08$ year$^{-1}$ is 0.04 year$^{-1}$.

The r-mode frequencies, the peak powers and the amplitudes for Mt Wilson data (Sturrock & Bertello 2010) and MDI data are listed in Tables 1 and 2, respectively. There is broad agreement, although r modes show up more prominently in Mt Wilson data, due perhaps to the dataset being of longer duration.

4. SIGNIFICANCE TEST

A significance analysis must take into account the possible interplay between the several frequencies contributing to the CPS. Accordingly, we have carried out a significance estimate as follows. Noting that, for the sidereal frequency 12.08 year$^{-1}$, the nine r modes for $l = 2,\ldots,10$, $m = 1$ all fall in the frequency band $\nu = 0 - 5$ year$^{-1}$, we have randomly selected nine frequencies, uniformly distributed in this band, and computed the corresponding CPS, $G(U)$. This procedure was repeated 100,000 times. The resulting distribution of CPS values is shown, as a logarithmic display, in Figure 5. For only 81 out of the 100,000 trials is the CPS value as large as or larger than the actual value (230) found in Figure 3. We infer that the group of r-mode oscillations with $\nu_R = 12.08$ year$^{-1}$, shown in Figure 4, has a significance level of approximately 0.1%.

However, the value $\nu_R = 12.08$ year$^{-1}$ does not give the best *frequency* fit (in the sense of the least rms frequency discrepancy $\sigma$): that is obtained with $\nu_R = 12.34$ year$^{-1}$, yielding $\sigma_{min} = 0.03$ year$^{-1}$. The identification of the modes in this fit is unchanged.

We have also examined the possibility that some r modes with $m > 1$ might alternatively account for some of the peaks in the power spectrum since (allowing for $m > 1$) the mode identifications are not unambiguous. For instance, the $l = 3, m = 2$ mode has the same frequency as the $l = 2, m = 1$ mode, and the $l = 4, m = 2$ mode has the same frequency as the $l = 5, m = 3$ mode. To obtain an overall assessment of the significance of the modes with $l = 2,...,10, m = 2$, we have analyzed this sequence in the same way that we analyzed the $l = 2,...,10, m = 1$ sequence. We find that, for $\nu_R = 12.08$, the CPS value is only 80, indicating that modes with $m = 2$ alone fare significantly worse than those with $m = 1$.

However, if we focus on the *frequencies* rather than the CPS values, we arrive at a somewhat different conclusion. If we admit modes with *both* m = 1 and m = 2 and again restrict ourselves to $l = 2,...,10$, we find that the best fit is for $\nu_R = 12.22$ year$^{-1}$, yielding $\sigma_{min} = 0.03$ year$^{-1}$. The identification of the two modes with the lowest frequencies are now $(l,m) = (9,1)$ and $(8,2)$; the other two modes remain the same as when $m = 1$ was imposed. A Monte Carlo simulation conducted by replacing the observed frequencies with four random frequencies, uniformly distributed over $0 - 5$ year$^{-1}$, yields a 2% probability of obtaining by chance a value of $\sigma$ no greater than $\sigma_{min}$. If all the modes with $l \leq 10$ are admitted, then the best fit is found for $\nu_R = 12.24$ year$^{-1}$ with $\sigma_{min}$ a little smaller than before. There is now a probability of 25% of obtaining at least as good a fit by chance.

Since it was our analyses of Mt Wilson and LMSU data that led us to focus on modes with $m = 1$, it would be informative to re-investigate those datasets relaxing that restriction.

## 5. DISCUSSION

The results of our power-spectrum analysis of MDI radius measurements are fully consistent with earlier analyses of both Mt Wilson radius measurements and LMSU $^{90}$Sr beta-decay-rate measurements. As shown in Tables 1 and 2 of this article and in Table 2 of Sturrock et al. (2013), all three datasets are consistent with the presence of r modes in a region of the solar interior where the sidereal rotation rate is 12.08 year$^{-1}$. There is an especially good match between the expected r-mode frequencies and those of oscillations in the Mt Wilson data. This closer match may be attributed to the fact that this dataset is the longest (length 32.31 years compared with 13.44 years for MDI and 6.51 years for LMSU).

However, the Mt Wilson Observatory has the capability of processing all of its observations to derive radius measurements from the beginning of 1970 to January 22, 2013, and we plan to carry out an analysis of the complete Mt Wilson dataset when that is available. This analysis should yield more extensive information about solar r-mode oscillations, including the time-dependence of these oscillations. We also plan to carry out a comparative analysis of Mt Wilson and MDI radius measurements for the duration of MDI observations (1997.843 to 2011.282, for which the Mt Wilson radius measurements are already available). A comparative analysis for the common time interval will be especially useful, since the analysis of beta-decay data indicates that solar oscillations tend to be intermittent and to drift in frequency. The results of these analyses will be published at a later date.

**Our current results are** compatible with a model of the Sun with a slowly rotating core, as is suggested by the power spectrum analysis of Super-Kamiokande solar neutrino data, which has a peak at 9.43 year$^{-1}$ (Sturrock & Scargle 2006). If this oscillation is interpreted as the nonaxisymmetric modulation of the neutrino flux by the RSFP (Resonant Spin Flavor Precession: Akhmedov 1988; Lim & Marciano 1988) process caused by a nonaxisymmetric magnetic field, it implies a sidereal rotation rate of 10.43 year$^{-1}$. This in turn suggests that there may be an inner tachocline separating the slowly rotating core from the radiative envelope which rotates at about 13.8 year$^{-1}$ sidereal. The median rotation rate would be about 12.1 year$^{-1}$, compatible with the value 12.08 year$^{-1}$ inferred from our r-mode analysis.

We postulate that the r modes interact with an internal magnetic field. If that field has predominantly an $m = 1$ structure, as would have arisen from the decay of an initially random field over the main-sequence life of the Sun, this would favor the generation of r modes with $m = 1$, for all values of $l$, consistent with our findings for the LMSU $^{90}$Sr data, Mt Wilson data, and MDI data. (By contrast, r-mode oscillations related to solar activity are not restricted to $m = 1$ values; Sturrock et al. 2013.) A magnetic structure is likely to influence the balance of forces that could lead to a change in radius and influence the neutrino flux. This model also appears to be consistent with the expectation that r-mode oscillations are unstable, and can therefore grow to large amplitudes, in a region where there is a significant radial gradient in the rotation velocity (Papaloizou & Pringle, 1978; Sturrock et al. 2013).

There is as yet no consensus concerning the core rotation frequency as derived from helioseismologic data inversion (Howe, 2009). Chaplin et al. (2004) have demonstrated that, with the available data, a rotational frequency difference of less than 3.5 year$^{-1}$ could not be detected at the $1\sigma$ level. On the other hand, model fitting by Elsworth et al. (1995) (which can be more precise, albeit risking lower accuracy from potentially less reliable assumptions in defining the model) suggests that the solar core rotates significantly more slowly than the envelope. This conclusion is in accord with our conjectured interpretation of radius data, and is also in accord with Super-Kamiokande solar neutrino data that exhibits an oscillation of frequency 9.43 year$^{-1}$ (Sturrock et al. 2005a), indicative of rotation of a solar core with a sidereal rotation frequency of 10.43 year$^{-1}$.

Finally, we inquire into the origin of the remaining peaks in the power spectrum of Figures 2 and 4, asking whether they too can be the result of low-degree r modes with azimuthal orders different from unity. We have accordingly carried out a least-squares fit of theoretical eigenfrequencies of all modes with $l \leq 10$ to the 15 most prominent peaks in Figure 4. The best fit is found for $\nu_R = 12.34$ year$^{-1}$, yielding $\sigma_{\min} = 0.04$ year$^{-1}$, but with a probability of 40% of obtaining such a result with randomly distributed frequencies. Hence we cannot say whether or not it is likely that the remaining peaks are due to r modes.

There is a need for detailed modeling to determine whether r modes, together with an appropriate magnetic structure, can indeed explain both the magnitude of the change in radius and the magnitude of the postulated change in neutrino flux. More generally, we draw attention to the prospect that solar observations may help clarify the apparent variability of certain beta-decay rates and that, conversely, the further analysis of beta-decay rates may give new information concerning solar structure and solar processes.

We express our appreciation to Marcelo Emilio for help in processing the MDI radius measurements, and to Ephraim Fischbach and Alexander Kosovichev for helpful discussions related to this project. DOG thanks Philip H. Scherrer for his hospitality. This work was supported in part by NASA Contract NAS5-02139.

Table 1 Comparison of peaks in Mt Wilson power spectrum with those expected of r-modes for m = 1 and a sidereal rotation frequency of 12.08 year$^{-1}$.

| $l$ | Calculated Frequency (year$^{-1}$) | Mt Wilson Peak Frequency | Mt Wilson Peak Power | Mt Wilson Power at Calculated Frequency | Mt Wilson Peak Amplitude mas |
|---|---|---|---|---|---|
| 2 | 4.03 | 4.04 | 69.5 | 69.1 | 2.14 |
| 3 | 2.01 | 2.01 | 245.6 | 245.6 | 3.96 |
| 4 | 1.21 | 1.21 | 497.5 | 497.5 | 5.70 |
| 5 | 0.81 | 0.80 | 253.0 | 234.4 | 4.07 |
| 6 | 0.58 | 0.57 | 405.1 | 359.2 | 5.16 |
| 7 | 0.43 | 0.42 | 315.5 | 278.4 | 4.54 |
| 8 | 0.34 | 0.35 | 126.8 | 110.1 | 2.91 |
| 9 | 0.27 | | | | |
| 10 | 0.22 | 0.22 | 541.9 | 541.9 | 5.94 |

Table 2 Comparison of peaks in MDI power spectra with those expected of r-modes for m = 1 and a sidereal rotation frequency of 12.08 year$^{-1}$.

| $l$ | Calculated Frequency (year$^{-1}$) | MDI Peak Frequency | MDI Peak Power | MDI Power at Calculated Frequency | MDI Amplitude (mas) |
|---|---|---|---|---|---|
| 2 | 4.03 | | | | |
| 3 | 2.01 | 2.07 | 65.5 | 58.0 | 5.1 |
| 4 | 1.21 | 1.20 | 30.8 | 30.4 | 3.4 |
| 5 | 0.81 | | | | |
| 6 | 0.58 | 0.62 | 66.4 | 40.9 | 4.9 |
| 7 | 0.43 | | | | |
| 8 | 0.34 | | | | |
| 9 | 0.27 | | | | |
| 10 | 0.22 | 0.25 | 82.7 | 55.1 | 1.7 |

FIGURES

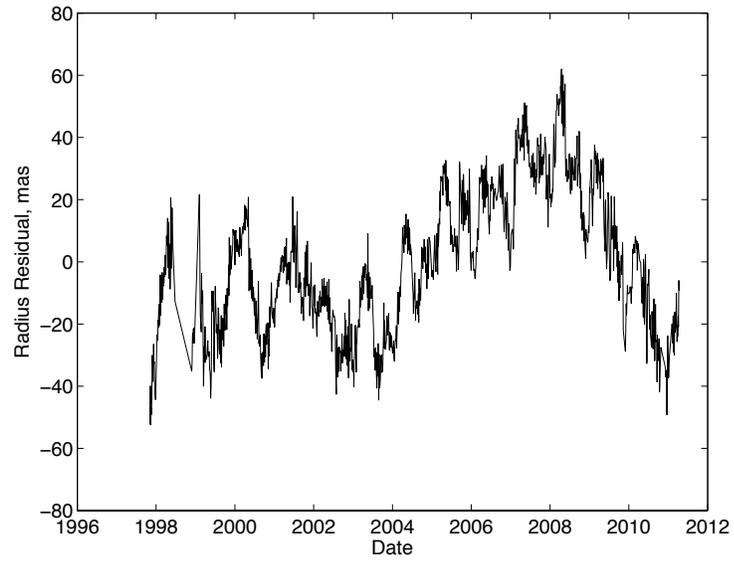

Figure 1. MDI binned radius residuals in milli-arc-seconds. The standard deviation of the residuals is 22.5 milli-arc-seconds.

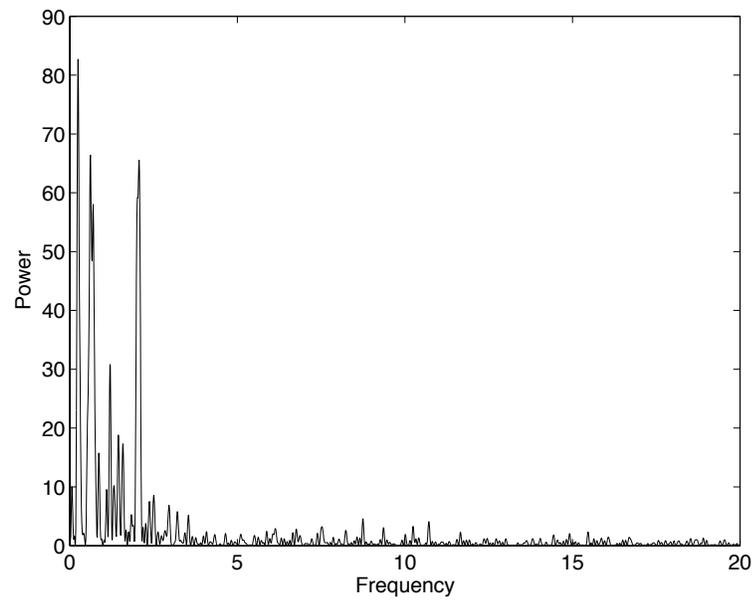

Figure 2. Power spectrum formed from MDI semi-diameter data after removing oscillations with frequencies 0.08 year$^{-1}$ and 0.15 year$^{-1}$ and an annual oscillation.

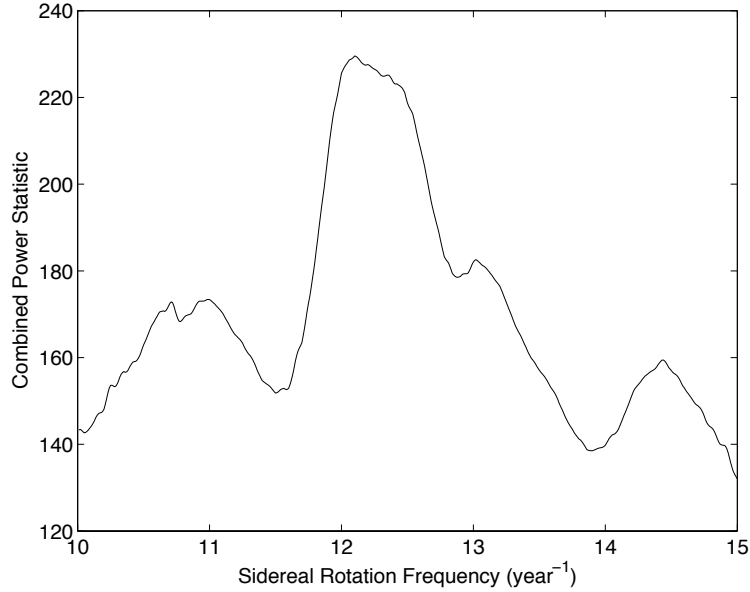

Figure 3. Combined power statistic formed from the values of the power (after removing oscillations with frequencies 0.08 year$^{-1}$ and 0.25 year$^{-1}$ and an annual oscillation, as in Figure 2) at the r-mode frequencies corresponding to $m=1, l=2,...,10$ plotted as a function of the sidereal rotation frequency $\nu_R$.

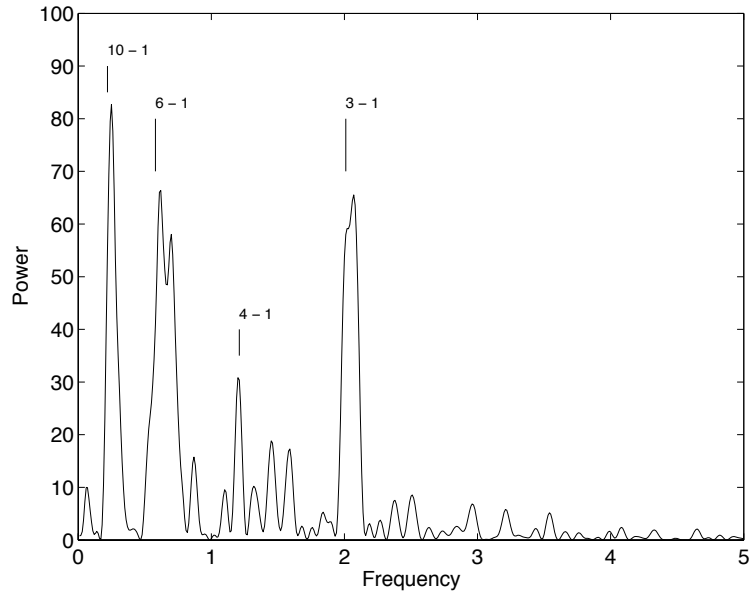

Figure 4. Low-frequency range of the power spectrum formed from MDI data after removing oscillations with frequencies 0.08 year$^{-1}$ and 0.15 year$^{-1}$ and an annual oscillation, identifying four r-mode oscillations (by their $l$ and $m$ values) corresponding to a sidereal rotation frequency of 12.08 year$^{-1}$.

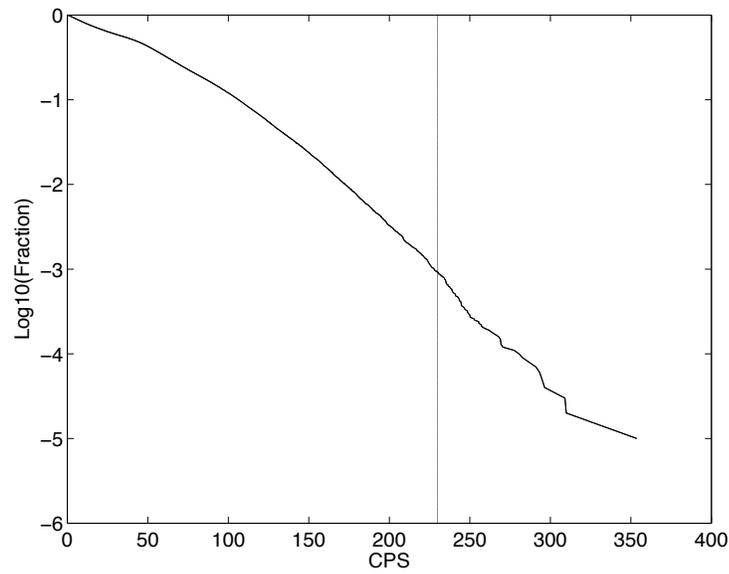

Figure 5. Outcome of 100,000 Monte Carlo simulations of the combined power statistic (CPS) computed from 9 random frequencies in the range 0 – 5 year$^{-1}$ applied to the power spectrum of the semi-diameter measurements illustrated in Figure 4. Plotted is the fraction of trials with CPS exceeding the value of the abscissa. The vertical line marks the value of the CPS (230) computed from the frequencies given by Equation (1) with $\nu_R = 12.08$ year$^{-1}$ and $l = 2,3,...,10$, $m = 1$. Only 0.08% of the simulations have CPS as large as or larger than that value.